\documentclass[preliminary,copyright,creativecommons]{eptcs}
\providecommand{\event}{LSFA 2023} % Name of the event you are submitting to

\usepackage{iftex}

\usepackage{booktabs}   %% For formal tables:
\usepackage{subcaption} %% For complex figures with subfigures/subcaptions
\usepackage[T1]{fontenc}
\usepackage{hyperref}
\usepackage{amsfonts,amsmath,amssymb,amstext,amsthm}
\newtheorem{theorem}{Theorem}[section]
\usepackage{enumitem}
\usepackage{graphicx}
\usepackage{xcolor}
\usepackage{listings}
\usepackage{multicol}
\usepackage{pdflscape}
\usepackage{csquotes}

\renewcommand{\gcd}{{\texttt{gcd}}}
\newcommand{\ndiv}{{\texttt{div}}}
\newcommand{\rem}{{\texttt{rem}}}
\newcommand{\fsI}{{\texttt{fsI}}}
\usepackage[misc]{ifsym}
\usepackage{fontawesome}
\newcommand{\ExtLink}{{ \color{blue}\faExternalLink}}

\newcommand{\linkalgebraNasalib}[1]{\href{https://github.com/nasa/pvslib/tree/master/algebra}{{\color{cyan}#1}\ExtLink}}

\newcommand{\linkThalgebra}[4]{\href{https://github.com/nasa/pvslib/tree/master/algebra/#1\#L#2-L#3}{{\color{cyan}#4}\ExtLink}}
\newcommand{\linkThalgebraex}[4]{\href{https://github.com/nasa/pvslib/tree/master/algebra/algebra_examples/#1\#L#2-L#3}{{\color{cyan}#4}\ExtLink}}

\title{Formalizing Factorization on Euclidean Domains and Abstract Euclidean Algorithms\thanks{Project supported by FAPDF DE 00193.00001175/21-11, CNPq Universal  409003/21-2, and FAPEG 202310267000223 grants. Last author partially funded by CNPq grant 313290/21-0.}} 
\author{Thaynara Arielly de Lima
\institute{Universidade Federal de Goiás, Brasil}
\email{thaynaradelima@ufg.br}
\and
Andréia {Borges Avelar} 
\institute{Universidade de Brasília, Brasil}
\email{andreiaavelar@unb.br} 
\and 
André Luiz {Galdino}
\institute{Universidade Federal de Catalão, Brasil}
\email{andregaldino@ufcat.edu.br}
\and 
Mauricio {Ayala-Rincón}
\institute{Universidade de Brasília, Brasil}
\email{ayala@unb.br}
}

\newcommand{\titlerunning}{Formalizing Factorization on Euclidean Domains} 
\newcommand{\authorrunning}{T.A. de Lima, A.B. Avelar, A.L. Galdino and M. Ayala-Rincón}

\hypersetup{
  bookmarksnumbered,
  pdftitle    = {\titlerunning},
  pdfauthor   = {\authorrunning},
  pdfsubject  = {EPTCS},               % Consider adding a more appropriate subject or description
}

\begin{document}

\maketitle

\begin{abstract}
This paper discusses the extension of the Prototype Verification System (PVS) sub-theory for rings, part of the PVS {\tt algebra} theory, with theorems related to the division algorithm for Euclidean rings and Unique Factorization Domains that are general structures where an analog of the Fundamental Theorem of Arithmetic holds. First, we formalize the general abstract notions of divisibility, prime, and irreducible elements in commutative rings, essential to deal with unique factorization domains. Then, we formalize the landmark theorem, establishing that every principal ideal domain is a unique factorization domain. Finally, we specify the theory of Euclidean domains and formally verify that the rings of integers, the Gaussian integers, and arbitrary fields are Euclidean domains. 
To highlight the benefits of such a general abstract discipline of formalization, we specify a Euclidean gcd algorithm for Euclidean domains and formalize its correctness. Also, we show how this correctness is inherited under adequate parameterizations for the structures of integers and Gaussian integers.
\end{abstract}

\section{Introduction}
The NASA PVS {\tt algebra} library  (\cite{PVSalgebra}) was recently enriched with a series of theorems related to the theory of rings. The extension includes complete formalizations of the isomorphism theorems for rings, principal, prime, and maximal ideals, and a general abstract version of the Chinese Remainder Theorem (CRT), which holds for abstract rings, including non-commutative rings. The benefit of formalizing algebraic results from this abstract theoretical perspective was made evident by showing how, from the abstract version of CRT, the well-known numerical version of CRT for the ring of integers $\mathbb{Z}$ was formalized  \cite{LimaGAA21}.

In this work, we give another substantial step towards enriching the PVS abstract algebra library by formalizing properties about factorization in commutative rings regarding both unique factorization domains and Euclidean rings. 
Roughly, unique factorization domains are abstract structures for which a general version of the Fundamental Theorem of Arithmetic holds. On the other hand, Euclidean rings are equipped with a norm that allows defining a suitable generalization of Euclid's division lemma and, consequently, of notions such as the greatest common divisor (\gcd). The practicality of \gcd\ is well-known in the ring $\mathbb{Z}$. Nevertheless, mathematicians know this notion is of fundamental importance in abstract Euclidean domains for which, in general, \gcd\ should and may be defined in different ways. 

\begin{figure*}
\includegraphics[width=1.021\textwidth]{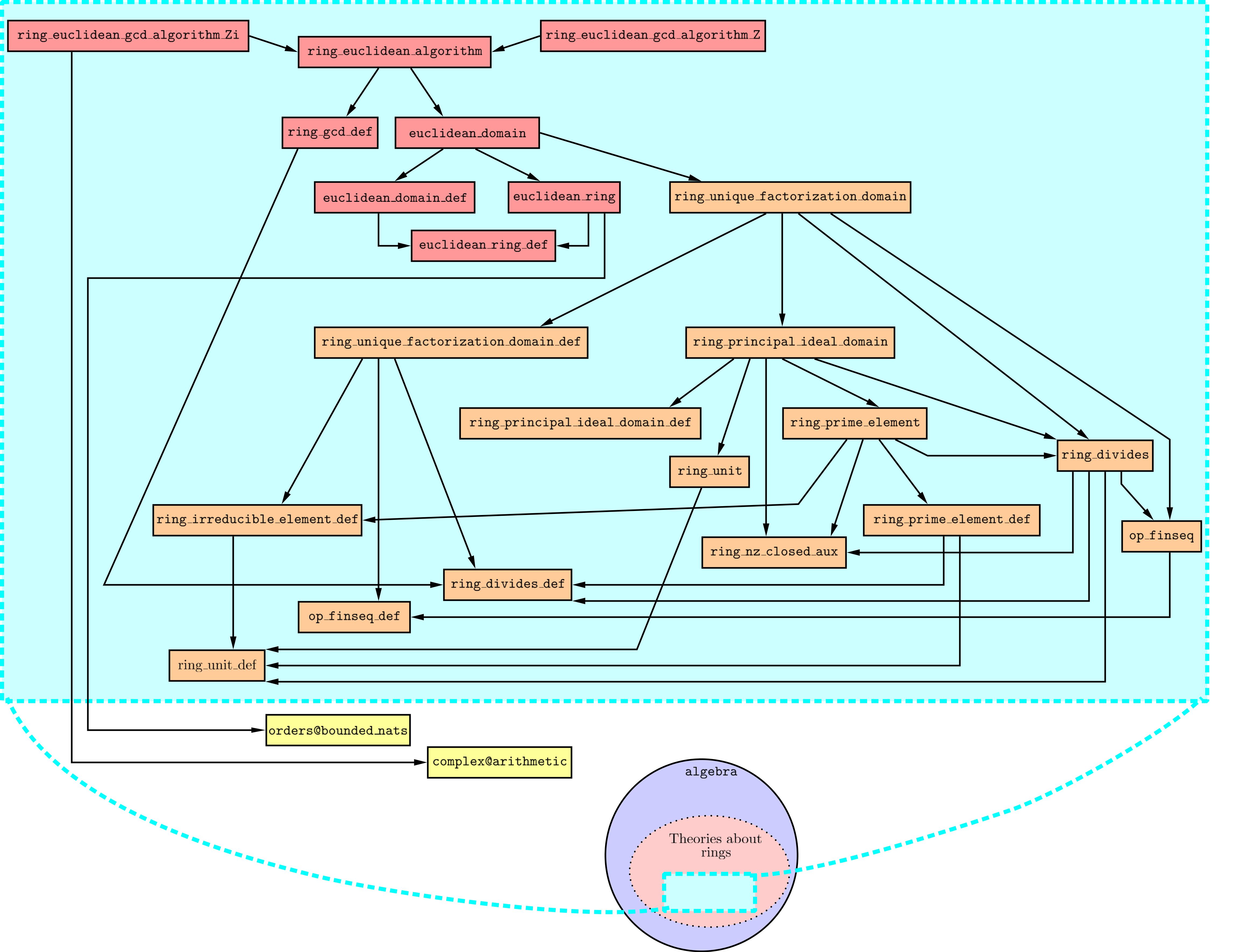} 
\caption{Ring theories expanding the PVS {\tt algebra} library} \label{fig:hierarchy_UFD_Euclidean}
\end{figure*}

Figure \ref{fig:hierarchy_UFD_Euclidean}  highlights the subtheories subject of the extension to the PVS theory {\tt algebra} discussed in this paper. The red ones are related to Euclidean rings, and \gcd\ algorithms for Euclidean domains, and the orange ones are those related to unique factorization domains. The extension includes 210 new formulas enlarging the theory {\tt algebra} from 1356 (cf \cite{LimaGAA21}) to 1566 formalized lemmas.   

The primary motivation to formalize such structures is their potential theoretical and practical applications. Using the example of \gcd, one can provide a general abstract version of the Euclidean algorithm to determine a \gcd\ between two elements (Euclidean \gcd\ algorithm) in a Euclidean domain. Since the ring of integers $\mathbb{Z}$, the Gaussian integers $\mathbb{Z}[i]$ (which are the subset of complex numbers whose real and imaginary parts are integer numbers) and rings of polynomials over integral domains are particular Euclidean domain structures, the Euclidean \gcd\ algorithm can be applied over them, in a relatively straightforward manner, to compute \gcd s in different manners, not only for the structures mentioned above but for a variety of  Euclidean domains. 

Also, every element of a unique factorization domain can be factorized as a finite number of irreducible elements, and one can prove that Euclidean domains are unique factorization domains. These properties allow us to introduce modular arithmetic, verify generic versions of Euler's Theorem and Fermat's Little Theorem for Euclidean domains, and promote factorization in Euclidean domains as a convenient feature to develop efficient algorithms in symbolic computation \cite{Lichtblau13}, \cite{EderPP17}. Thus, formalizing the main results about unique factorization and Euclidean domains would allow the formal verification of more complex theories involving such structures in their scope.

The main contributions of this paper are listed below. 

\begin{itemize}

\item We formalize the abstract notions of divisibility, prime, and irreducible elements in commutative rings, which are essential to deal with unique factorization domains. In integral domains, prime elements are irreducible. The converse is not true in general. Among other properties, we formalize the theorem that establishes that irreducible elements are also prime in principal ideal domains (as well-known, it holds in $\mathbb{Z}$).  

\item We specify unique factorization domains and formalize the theorem that every principal ideal domain is a unique factorization domain, which is a landmark result in abstract algebra. %and whose formalization depends on Zorn's lemma and constructing an ascending chain of ideals. 
\item We specify the notion of Euclidean domains and formally verify that the rings $\mathbb{Z}$ and  $\mathbb{Z}[i]$ and any arbitrary field are Euclidean domains.

\item We specify the general abstract notion of \gcd\ for commutative rings, providing a general Euclidean  \gcd\ algorithm for Euclidean domains, formalizing its correctness. Using this result, we parameterize the adequate norms and \gcd\ relations for the rings $\mathbb{Z}$ and  $\mathbb{Z}[i]$; thus, obtaining straightforwardly the correctness of such instantiations of the abstract algorithm for these Euclidean domains. In this manner, we illustrate the benefits of maintaining the abstract general discipline of formalization for algebraic theories and the potential of such a discipline for application in concrete algebraic structures. 
\end{itemize}

\noindent{\bf Organization of the paper}. Section \ref{sec:Euclidean_domains} presents a theoretical overview of unique factorization and Euclidean domains, pointing out the main concepts and results. Also, it comments on some differences between pen-and-paper proofs presented in Hungerford's textbook \cite{hungerford80} and this formalization. Section \ref{sec:GCD_algorithms} discusses the aspects of the formalization of the Euclidean \gcd\ Algorithm for Euclidean Domains, as well as its application for two particular cases. Section \ref{sec:related_work} discusses related work and work in progress.
Finally, Section \ref{sec:conclusion} concludes and suggests future work. The formalizations were developed using PVS and are available at \linkalgebraNasalib{algebra}.

\section{Formalization of Euclidean Domains} \label{sec:Euclidean_domains}
Notions such as prime element, division, and \gcd\ between two elements and some landmark results, including the Fundamental Theorem of Arithmetic, Euclid's division lemma, and Euclidean Algorithm, are well established and widespread for the ring of integers. Such concepts and general versions of interesting results are extended for abstract algebraic structures (\cite{hungerford80}, \cite{dummit03}, \cite{Fraleigh2003}) and are the scope of our formalization. 

This section gives a theoretical overview of the central notions and properties and discusses the PVS features used in their formalization. An excellent description of the semantics of PVS is available as \cite{owreshankar1999}. In addition, to highlight crucial differences between pen-and-paper vs formalized proofs, some analytical concepts and results are presented as enunciated in Chapter III of Hungerford's textbook \cite{hungerford80}.

\subsection{Prime and irreducible elements on rings}
The definitions of prime and irreducible elements rely on the general concept of divisibility on a ring. 
The specification of the notions of divisibility and associated elements are specified as the curried predicates given in Specification \ref{code:divides_def}. These predicates are abstracted for any ring structure given as their first argument, {\tt R}. In PVS, types are built from predicates; for example,  {\tt ring?} is used to build the type {(ring?)}.

\begin{lstlisting}[float=*, xleftmargin=2mm, linewidth=15.8cm, label = code:divides_def , caption = {Divisibility and associated elements in the sub-theory \linkThalgebra{ring_divides_def.pvs/}{26}{45}{\tt ring\_divides\_def}},  mathescape, frame = single]
divides?(R : (ring?))(a: (R - {zero}), b: (R)): bool =  EXISTS (x: (R)): a*x = b

associates?(R: (ring?))(a,b:(R - {zero})): bool = divides?(R)(a,b) AND 
                                                  divides?(R)(b,a)
 \end{lstlisting}
 In Hungerford's textbook, the definition of divisibility relies on a commutative ring. It avoids the discrimination between an element's left or right divisor, and since the main results demand a commutative ring in the hypothesis, it is a reasonable requirement.  
 However, commutativity is not a crucial property in such a notion since it only depends on the multiplication operation in a ring. Because of that, we opted to generalize the definition and specify divisibility on non-necessarily commutative rings as ({\tt divides?(R)(a,b)}).
 Another interesting remark is related to the specification of {\tt associates?(R)(a,b)}: Hungerford's textbook omits that the type of the parameters {\tt a} and {\tt b} are non-zero elements. Of course, this is obvious since it is required in the definition of {\tt divides?(R)(a,b)}. However, the lack of such a hypothesis is recurrent in several statements throughout the textbook that require it (for example, in Theorem \ref{theo:divides}). 
 
 In the sub-theory \linkThalgebra{ring_divides.pvs}{23}{87}{\tt ring\_divides}, we formalized the properties related to the divisibility stated in Theorem \ref{theo:divides}. Some of them involve the object ``unit''. In a ring $(R,+,*,zero,one)$ with multiplication identity $one$, an element $u$ is called a \textit{unit} if $u$ is left- and right-invertible; that is, if there exist elements $u_1^{-1}, u_2^{-1} \in R$ such that $u*u_1^{-1} = u_2^{-1}*u = one$.

 \begin{theorem}[Th.3.2, Hungerford \cite{hungerford80}] \label{theo:divides}
 Let $a, b$, and $u$ be elements of a commutative ring $R$ with identity. 
 \begin{enumerate}[label=(\roman*)]
     \item $a$ divides $b$ (denoted as $a\; |\; b$) if and only if $(b) \subset (a)$, where $(x)$ denotes the principal ideal generated by $x$. 
     \item $a$ and $b$ are associates if and only if $(a) = (b)$.
     \item $u$ is a unit if and only if $u\;|\; r$ for all $r \in R$. 
     \item $u$ is a unit if and only if $(u) = R$. 
     \item The relation ``$a$ and $b$ are associates'' is an equivalence relation on $R$.
     \item If $a = br$, where $r \in R$ is a unit, then $a$ and $b$ are associates. If $R$ is an integral domain, then the converse is true. 
     
 \end{enumerate}
 \end{theorem}

 Theorem \ref{theo:divides} has a straightforward formalization due to the robustness of the formal framework previously developed for rings and principal ideals \cite{LimaGAA21}. The formalization of the properties \textit{(i)}, \textit{(ii)}, and  \textit{(iv)} illustrates it clearly. In fact, by definition, $(a)$ denotes the intersection of all ideals in $R$ containing the element $a$. The lemma \linkThalgebra{ring_principal_ideal.pvs}{48}{50}{\tt principal\_ideal\_charac} in theory {\tt ring\_principal\_ideal} characterizes  $(a)$ as the set \linkThalgebra{ring_one_generator.pvs}{50}{52}{\tt one\_gen(R)(a)} in the theory {\tt ring\_one\_generator}. 
The last characterization depends on a sum, specified as {\tt R\_sigma}, over elements of a function in the ring $R$, defined over abstract types, as given in the theory \linkThalgebra{ring_basic_properties.pvs}{25}{226}{\tt ring\_basic\_properties}. The constructor {\tt R\_sigma} generalizes constructors in the NASA PVS library (nasalib) built for specific theories as the theory of reals.  
Also, since $R$ is a commutative ring with identity, the lemma 
\linkThalgebra{ring_with_id_one_generator.pvs}{55}{57}{\tt commutative\_id\_one\_gen\_charac} provides a much simpler characterization of the set {\tt one\_gen(R)(a)}; indeed, such characterization simplifies the analysis of properties  \textit{(i)},  \textit{(ii)}, and  \textit{(iv)} since $(a)$ can be built as the set $aR = \{ar \;:\;  r \in R\}$.

 From the concepts of divisibility and unit, we specified prime and irreducible elements on a ring with identity as the predicates given in the  Specification \ref{code:prime_irreducible_def}.
 
 \begin{lstlisting}[float=*, xleftmargin=2mm, linewidth=15.8cm, label = code:prime_irreducible_def , caption = {Irreducible and prime elements in the subtheories \linkThalgebra{ring\_irreducible\_element\_def.pvs}{22}{45}{\tt ring\_irreducible\_element\_def} and \linkThalgebra{ring_prime_element_def.pvs}{22}{38}{\tt ring\_prime\_element\_def}, respectively},  mathescape, frame = single]
R_irreducible_element?(R : (ring_with_one?))(x:(R)): bool = x/=zero AND 
    (NOT unit?(R)(x)) AND
    (FORALL (a,b:(R)): x = a*b IMPLIES (unit?(R)(a) OR unit?(R)(b)))

R_prime_element?(R : (ring_with_one?))(x:(R)): bool = x/=zero AND (NOT unit?(R)(x)) AND
 (FORALL (a,b:(R)): divides?(R)(x, a*b) IMPLIES   
                    divides?(R)(x, a) OR divides?(R)(x, b))
 \end{lstlisting}

In the ring of integers,  prime and irreducible elements are indistinguishable. However, this is not true for all rings. For instance, $2$ is prime but not irreducible in $\mathbb{Z}_6$. Theorem \ref{theo:prime_irreducible_prop} gives some properties regarding prime and irreducible elements formalized in the subtheories \linkThalgebra{ring_principal_ideal_domain.pvs}{24}{118}{\tt ring\_principal\_ideal\_domain} and \linkThalgebra{ring_prime_and_irreducible_element.pvs}{23}{117}{\tt ring\_prime\_and\_irreducible\_element}. It shows, among other results, that prime and irreducible elements are equal over principal ideal domains.
 
 \begin{theorem}[Th.3.4, Hungerford \cite{hungerford80}] \label{theo:prime_irreducible_prop}
 Let $p$ and $c$ be nonzero elements in an integral domain $R$. 
 \begin{enumerate}[label=(\roman*)]
     \item $p$ is prime if and only if $(p)$ is a nonzero prime ideal;
     \item $c$ is irreducible if and only if $(c)$ is maximal in the set $S$ of all proper principal ideals of $R$.
     \item Every prime element of $R$ is irreducible. 
     \item If $R$ is a principal ideal domain, then $p$ is prime if and only if $p$ is irreducible.
     \item Every associate of an irreducible [resp. prime] element of $R$ is irreducible [resp. prime].
     \item The only divisors of an irreducible element of $R$ are its associates and the units of $R$.
 \end{enumerate}
  \end{theorem}
Although the result is stated for integral domains, Hungerford advises that a weakened hypothesis can be considered in some parts of the theorem. We formalize the results using the minimum number of required conditions and detect that items (i) and (vi) of the Theorem \ref{theo:prime_irreducible_prop} hold for commutative rings with identity. 

Properties \textit{(i)}, \textit{(ii)}, and \textit{(iii)} form the basis for the formalization of the 
characterization of primes as irreducible elements over principal ideal domains, given in property \textit{(iv)} and specified as the lemma \linkThalgebra{ring_principal_ideal_domain.pvs}{56}{57}{\tt PID\_prime\_el\_iff\_irreducible}. 
The sufficiency of the property \textit{(iv)}, established in the property \textit{(iii)}, is verified as the lemma \linkThalgebra{ring_prime_and_irreducible_element.pvs}{60}{61}{\tt prime\_el\_is\_irreducible}. It follows in a relatively straightforward manner from the definition of prime elements and the result that the multiplicative cancelation law holds for non-zero elements in integral domains (lemma \linkThalgebra{ring_nz_closed_aux.pvs}{43}{44}{\tt nzd\_R\_cancel\_left}) since integral domains have no zero divisors. 

On the other hand, the necessity of \textit{(iv)} is trickier since it depends on properties \textit{(i)}, \textit{(ii)}, and other additional previous results developed for rings with identity and maximal ideals. 
 Property \textit{(i)} is specified by the lemma \linkThalgebra{ring_prime_and_irreducible_element.pvs}{48}{50}{\tt prime\_el\_iff\_prime\_ideal}. Its proof depends on the lemmas \linkThalgebra{ring_prime_ideal.pvs}{41}{50}{{\tt prime\_ideal\_prop1} and {\tt prime\_ideal\_prop2}} formalized in theory {\tt ring\_prime\_ideal}, which provide a characterization of prime ideals over commutative rings.
Lemma \linkThalgebra{ring_prime_and_irreducible_element.pvs}{53}{57}{\tt el\_irred\_iff\_one\_gen\_maximal} specifies property \textit{(ii)}, which establishes that the principal ideal generated by an irreducible element is a maximal element in the set $S$ of all proper principal ideals of a ring. 
It is important to stress here that in the pen-and-paper proof of property \textit{(iv)} given in \cite{hungerford80}, Hungerford assumes the vital result that maximal elements in the previously mentioned set $S$ are maximal ideals in $R$. We formalized this property without this assumption as the lemma \linkThalgebra{ring_principal_ideal_domain.pvs}{50}{54}{\tt el\_max\_iff\_one\_gen\_maximal} in the sub-theory {\tt ring\_principal\_ideal\_domain}. 
Finally, the necessity of property \textit{(iv)} is concluded as follows. If $p$ is an irreducible element, then $(p)$ is a maximal element, according to {\tt el\_max\_iff\_one\_gen\_maximal}. Since $R$ is a ring with identity, $R^2 = R$ by lemma \linkThalgebra{ring_with_one_basic_properties.pvs}{63}{64}{\tt ring\_w\_one\_is\_idempotent}, which is formalized in the sub-theory {\tt ring\_with\_one\_basic\_properties}. Consequently, $(p)$ is a prime ideal by the lemma \linkThalgebra{ring_maximal_ideal.pvs}{39}{41}{\tt maximal\_prime\_ideal} and, by property (i), $p$ is a prime element.

%%%%%%%%%%%%%%%%%%%%%%%%%%%%%%%  
\subsection{Unique Factorization Domains} \label{sec:UFD} 
The well-known Fundamental Theorem of Arithmetic for integers states that any positive integer greater than $1$ can be factorized as a unique product of primes up to a permutation of such factors. Unique Factorization Domains (UFDs) are integral domains with an analogous theorem. The Specification \ref{code:UFD_def} shows the definition of UFDs. It depends on a sequence of irreducible elements {\tt fsIr?}$(R)(\fsI)$ on a ring $R$ with identity and a recursive operator {\tt op\_fseq}$(\fsI)$, as specified in the sub-theory \linkThalgebra{op_finseq_monoid_def.pvs}{22}{35}{\tt op\_finseq\_monoid\_def}, which multiplies the elements of such a sequence. The operator {\tt op\_fseq}$(\fsI)$ is specified over an abstract structure $(T,*,one)$  equipped with a binary operation $*$ and a constant $one$. 

From the point of view of formalization, such a general specification is very useful for two reasons: firstly, it allows the use of the operator {\tt op\_fseq}$(\fsI)$ in a variety of abstract and concrete structures (monoids, monads, groups, rings, integers, reals) by only adequately parameterizing the sub-theory {\tt op\_finseq\_def}; secondly, it avoids proof obligations, called in PVS \emph{ Type Correctness Conditions (TCCs)}, automatically generated by the system, since the operator is defined for elements of an abstract type, which provides more automation in our formal verification. Indeed, suppose such an operator was defined over elements of an algebraic structure, for example, a monad. To each application of that definition in a specific context, PVS will automatically generate a proof obligation to verify that {\tt op\_fseq}$(\fsI)$ acts on a sequence whose elements belong to a monad. This specification design would make the theory verification more onerous. It is advantageous to use polymorphism to formalize concepts and properties that hold for a non-interpreted type since it allows the reuse of such results in multiple contexts.

\begin{lstlisting}[float=*, xleftmargin=2mm, linewidth=15.8cm, label = code:UFD_def, caption = {Theory   \linkThalgebra{ring_unique_factorization_domain_def.pvs}{22}{43}{\tt ring\_unique\_factorization\_domain\_def} with the definition of unique factorization domain},  mathescape, frame = single]
fsIr?(R)(fsI: finseq[(R)]): bool = FORALL (i: below[length(fsI)]):
     R_irreducible_element?(R)(fsI(i))
   
unique_factorization_domain?(R): bool = integral_domain_w_one?(R) AND 
FORALL(a: (R)): a /= zero AND NOT unit?(R)(a) IMPLIES
 EXISTS(fsI:(fsIr?(R))):a = op_fseq(fsI) AND
 FORALL(fsIp:fsIr(R)):a = op_fseq(fsIp) IMPLIES length(fsI) = length(fsIp) AND
 EXISTS(phi:[below[length(fsI)]->below[length(fsI)]]): (bijective?(phi)) AND 
 FORALL(i:below[length(fsI)]): associates?(R)(fsIp(phi(i)),fsI(i)) 

\end{lstlisting}
In sub-theory   \linkThalgebra{ring_unique_factorization_domain.pvs}{23}{53}{\tt ring\_unique\_factorization\_domain}, we formalized the Theorem \ref{theo:principal_domain_UFD}, which is
a landmark result about UFDs. 

\begin{theorem}[Th.3.7, Hungerford \cite{hungerford80}]\label{theo:principal_domain_UFD}
Every principal ideal domain is a unique factorization domain. 
\end{theorem}

The formalization of the Theorem \ref{theo:principal_domain_UFD} has two main steps. We briefly comment on them.

\vspace{1mm}

\textbf{Step 1 -  Existence of a factorization}
\vspace{1mm}

First, previous subtheories established in the PVS theory {\tt algebra} were enriched with auxiliary results. 
The new lemma \linkThalgebra{ring_ideal.pvs}{106}{108}{{\tt chain\_ideal\_} {\tt union\_ideal}}, which states that the union of a chain of ideals in a ring $R$ is an ideal, is included in the sub-theory \linkThalgebra{ring_ideal.pvs}{24}{111}{\tt ring\_ideal}. The new lemma \linkThalgebra{ring_with_one_maximal_ideal.pvs}{75}{77}{{\tt nonzero\_} {\tt ring\_} {\tt exists\_maximal\_ideal\_aux}}, which proves that every ideal in a ring $R$ with identity, except $R$ itself, is contained in a maximal ideal in $R$, is added to the sub-theory \linkThalgebra{ring_with_one_maximal_ideal.pvs}{25}{83}{{\tt ring\_with\_} {\tt one\_maximal\_ideal}}. 

The formalization of this lemma considers an ideal $A \neq R$, $S = \{ B \subset R; \; B \mbox{ is ideal in } R,$  $B\neq R \mbox{ and } A \subset B \}$ and $\mathcal{C} = \{C_i \;|\; i\in I\}$ an arbitrary chain of ideals in $S$. 
We prove that the ideal $C = \bigcup C_i$ is an upper bound of the chain $\mathcal{C}$ in $S$ and, by using Zorn's lemma (available in the NASA PVS theory {\tt orders}), we conclude that $S$ has a maximal element, which is a maximal ideal in $R$. In the sub-theory \linkThalgebra{ring_principal_ideal.pvs}{24}{69}{\tt ring\_principal\_ideal}, we add the new lemma \linkThalgebra{ring_principal_ideal.pvs}{61}{66}{\tt stable\_chain}, which states that if $R$ is a principal ideal ring and $(a_1) \subset (a_2) \ldots$ is a chain of ideals in $R$, then for some positive integer $n$, $(a_j) = (a_n)$ for all $j \geq n$. The new lemma \linkThalgebra{ring_principal_ideal_domain.pvs}{59}{61}{\tt nonzero\_nonunit\_irreducible\_divides}, formalized in the sub-theory \linkThalgebra{ring_principal_ideal_domain.pvs}{24}{118}{\tt ring\_principal\_ideal\_domain}, states that every nonzero and non-unit element in a principal ideal domain is divided by an irreducible element. 

We conclude Step 1 by verifying that the subset of $R$ below, a principal ideal domain, is empty.

\begin{center}
{\tt non\_fact\_el\_set}$(R)$ =  $\left\{ \begin{array}{cl}x  \;:\; & x \mbox{ is a nonzero non-unit element} \\ 
&\mbox{in } R \mbox{ and cannot  be finitely} \\ 
&\mbox{factorized  into irreducible elements}
\end{array}\right\}$ 
\end{center}

In fact, if $a \in $ {\tt non\_fact\_el\_set}$(R)$, we could build an ascending chain of ideals, $(a) \subset (a_1) \subset\ldots$, contradicting the lemma {\tt stable\_chain}. The key to verifying it is to specify the recursive function \linkThalgebra{ring_principal_ideal_domain.pvs}{75}{79}{\tt phi$(n,R,a)$} showed in Specification \ref{code:function_chain}  (sub-theory \linkThalgebra{ring_principal_ideal_domain.pvs}{24}{118}{\tt ring\_principal\_ideal\_domain}) and verify that it is well defined whenever {\tt non\_fact\_el\_} {\tt set}$(R)$ is non-empty. In PVS, recursive definitions must be accompanied by a measure to prove termination. The measure has to decrease after each recursive step.
 \begin{lstlisting}[float=*, xleftmargin=2mm, linewidth=15.8cm, label = code:function_chain, caption = {Auxiliary function to build an ascending chain of ideals},  mathescape, frame = single]
 phi(n:nat, R:principal_ideal_domain, a:(non_fact_el_set(R))): 
  RECURSIVE (non_fact_el_set(R)) = 
   IF n = 0 then a
   ELSE  choose ({x : (non_fact_el_set(R))|
                 strict_subset?(one_gen(R)(phi(n-1, R, a)),one_gen(R)(x))})
   ENDIF  
  MEASURE n

 \end{lstlisting}

Whenever $a \in$  {\tt non\_fact\_el\_set}$(R)$, the choice of the element $a_1$, obtained by the function {\tt choose} in Specification \ref{code:function_chain}, is guaranteed. In fact, the lemma {\tt nonzero\_nonunit\_} {\tt irreducible\_divides}
ensures that $a = c a_1 $, where $c$ is irreducible. It implies that $a_1$ belongs to {\tt non\_fact\_el\_set}$(R)$ and satisfies the condition $(a) \subset (a_1)$ by Theorem \ref{theo:divides}\textit{(i)}.

% Given an element of $R$, the existence of a factorization is an intricate proof that depends on parts of theorems \ref{theo:divides} and \ref{theo:prime_irreducible_prop}, \tocheck{two other main lemmas regarding ideals} and Zorn's lemma that was previously formalized in the nasalib theory {\tt orders@zorn}. 

\vspace{1mm}

\textbf{Step 2: ``Uniqueness'' of a factorization}
\vspace{1mm}

By uniqueness we mean the existence of a bijective function between the elements of two factorizations mapping associated elements. 
First, we formalized the lemma \linkThalgebra{ring_prime_and_irreducible_element.pvs}{80}{84}{{\tt prime\_el\_} {\tt divides}} (sub-theory \linkThalgebra{ring_prime_and_irreducible_element.pvs}{23}{117}{\tt ring\_prime\_and\_irreducible\_element}) which states if a prime element $p$ in an integral domain divides the product $a_1 \ldots a_n$ then there exists $i$, $1 \leq i \leq n$, such that $p$ divides $a_i$. By \ref{theo:prime_irreducible_prop}\textit{(iii)}, 
$p$ is an irreducible element since $p$ is a prime element.
From this, if $a_1 \ldots a_n = a = b_1 \ldots b_m$, where $a_i, 1 \leq i \leq n$, and $b_j$, $1 \leq j \leq m$, are irreducible elements, then $a_1$ divides $b_j$, for some $j$. By Theorem \ref{theo:prime_irreducible_prop}\textit{(vi)}, $a_1$ and $b_j$ are associates. Using induction on $n$, we prove that $n = m$ and establish the required bijective function.

 %%%%%%%%%%%%%%%%%%%%%%%%%%%%%%%  
 \subsection{Euclidean Rings} \label{sec:Euclidean_rings} 
 
A Euclidean ring is a commutative ring $R$ equipped with a norm $\varphi$ over $R - \{zero\}$, where an abstract version of the well-known Euclid's division lemma holds. Euclidean rings and domains are specified in the subtheories \linkThalgebra{euclidean_ring_def.pvs}{24}{56}{\tt Euclidean\_ring\_def} and \linkThalgebra{euclidean_domain_def.pvs}{25}{38}{\tt Euclidean\_domain\_def} (Specification \ref{code:euclidean_ring_domain_def}). 
 
 \begin{lstlisting}[float=*, xleftmargin=2mm, linewidth=15.8cm, label = code:euclidean_ring_domain_def, caption = {Definitions of Euclidean rings and Euclidean domains},  mathescape, frame = single]
euclidean_ring?(R): bool = commutative_ring?(R) AND
EXISTS (phi: [(R - {zero}) -> nat]): FORALL(a,b: (R)):
  ((a*b /= zero IMPLIES phi(a) <= phi(a*b)) AND
   (b /= zero IMPLIES EXISTS(q,r:(R)): 
    (a = q*b+r AND (r = zero OR (r /= zero AND phi(r) < phi(b))))))
                                           

euclidean_domain?(R): bool = euclidean_ring?(R) AND integral_domain_w_one?(R)
\end{lstlisting}

% \begin{lstlisting}[float=*,label = code:euclidean_domain_def, caption = {Definition of Euclidean domains},  mathescape, frame = single]

% \end{lstlisting}

In sub-theory \linkThalgebra{euclidean_domain.pvs}{24}{50}{\tt Euclidean\_domain}, we formalized that elements of Euclidean ring can be factorized as irreducible elements by verifying Theorem \ref{theo:Euclidean_ring_principal_ideal}.
\begin{theorem}[Th.3.9, Hungerford \cite{hungerford80}] \label{theo:Euclidean_ring_principal_ideal}
A Euclidean ring $R$ is a principal ideal ring with identity. Consequently, every Euclidean domain is a unique factorization domain.
\end{theorem}
The verification makes use of the well-ordering principle over $\varphi(I^*) = \{\varphi(x) \in \mathbb{N}; \;  $x$ \; \in I - \{zero\}\}$, where $I$ is a nonzero ideal in $R$ and $\varphi$ is a norm on $R - \{zero\}$.  By choosing $a \in I$ such that $\varphi(a)$ is the minimum element of $\varphi(I^*)$, $b \in I$ satisfies $b = qa + r$, for some $q \in R$ and $r \in I$. From this, we infer that $r = 0$, since $r \neq 0$ contradicts the minimality of $\varphi(a)$. Consequently, $b = qa$ and $I \subset Ra \subset (a) \subset I$. The last guarantees that every ideal in $R$ is a principal ideal. By Theorem \ref{theo:principal_domain_UFD}, we have that a Euclidean principal ideal domain is a unique factorization domain.  

%{\color{red}In sub-theory }\linkThalgebraex{euclidean_domain.pvs}{24}{47}{\tt Euclidean\_domain} and \linkThalgebra{euclidean_domain.pvs}{24}{50}{}, 
We also formalized the results stating that the ring of integers (\linkThalgebraex{euclidean_domain.pvs}{44}{45}{integers\_is\_euclidean\_domain}) and any arbitrary field (\linkThalgebra{euclidean_domain.pvs}{44}{45}{field\_is\_euclidean\_domain}) are Euclidean domains. 
%The former is essential to verify the correctness of the \gcd\ algorithm for integers as an application of the abstract Euclidean algorithm for Euclidean domains, since it is a \emph{TCC} automatically generated by the system from the instantiation of the general version with the ring of integers. 

%%%%%%%%%%%%%%%%%%%%%%%%%%%%%%%  
\section{Formalization of \gcd\ Algorithm for Euclidean Domains} \label{sec:GCD_algorithms}

The theory \linkThalgebra{euclidean_ring_def.pvs}{24}{56}{\tt Euclidean\_ring\_def} includes two additional definitions to allow abstraction of acceptable Euclidean norms and associated functions fulfilling the properties of Euclidean rings (see Specification \ref{code:EuclideanRingFunctions}). 

The first definition is the relation \linkThalgebra{euclidean_ring_def.pvs}{40}{43}{\tt 
Euclidean\_pair?}. Given a Euclidean ring $R$ and a Euclidean norm of non-zero elements over the naturals $\phi : R\setminus \{zero\} \rightarrow \mathbb{N}$,  the predicate {\tt Euclidean\_pair?}$(R,\phi)$ holds whenever $\phi$ satisfies the constraints of a Euclidean norm over $R$.

The second definition is the curried relation given as \linkThalgebra{euclidean_ring_def.pvs}{47}{52}{\tt  Euclidean\_f\_phi?$(R,\phi)(f_\phi)$}. This relation holds whenever {\tt Euclidean\_pair?}$(R,\phi)$ holds, and $f_\phi$ is a function from $R\times R\setminus\{zero\}$ to  $R\times R$, such that for all pair of elements of $R$ in its domain, $f_\phi(a,b)$ gives a pair of elements, say $(div,rem)$ satisfying the constraints of Euclidean rings regarding the norm $\phi$: if $a \neq zero$, $a = div * b + rem$, and if $rem \neq zero$, $\phi(rem)< \phi(b)$.  These definitions are correct since the existence of such a $\phi$ and $f_\phi$ is guaranteed by the fact that $R$ is a Euclidean ring.  Also, notice that the decrement of the norm, i.e., $\phi(rem)< \phi(b)$, is the key to building an abstract Euclidean terminating procedure.

\begin{lstlisting}[float=*, xleftmargin=2mm, linewidth=15.8cm, label = code:EuclideanRingFunctions , caption = {Additional definitions in the sub-theory {\tt Euclidean\_ring\_def}},  mathescape, frame = single]
Euclidean_pair?(R : (Euclidean_ring?), phi: [(R - {zero}) -> nat]) : bool =
    FORALL(a,b: (R)): ((a*b /= zero IMPLIES phi(a) <= phi(a*b)) AND
                       (b /= zero IMPLIES 
                         EXISTS(q,r:(R)): (a = q*b+r AND 
                             (r = zero OR (r /= zero AND phi(r) < phi(b))))))

Euclidean_f_phi?(R : (Euclidean_ring?),
                 phi : [(R - {zero}) -> nat] | Euclidean_pair?(R,phi))
                (f_phi : [(R) , (R - {zero}) -> [(R),(R)]]) : bool = 
                 FORALL (a : (R), b :(R - {zero})): 
                  IF a = zero THEN f_phi(a,b) = (zero, zero)  
                  ELSE LET div = f_phi(a,b)`1, rem = f_phi(a,b)`2 IN
                     a = div * b + rem AND 
                    (rem = zero OR (rem /= zero AND phi(rem) < phi(b)))
                  ENDIF  
\end{lstlisting}

By using the previous two relations, a general abstract recursive Euclidean \gcd\ algorithm is specified in the sub-theory \linkThalgebra{ring_euclidean_algorithm.pvs}{23}{65}{\tt ring\_euclidean\_algorithm} as the curried definition  \linkThalgebra{ring_euclidean_algorithm.pvs}{36}{48}{\tt Euclidean\_gcd\_algorithm} (See Specification \ref{code:EuclideanAlgorithm}). The types of its arguments guarantee the correctness of this algorithm. Indeed, since allowed arguments $R, \phi$, and $f_\phi$ should satisfy {\tt  Euclidean\_f\_phi?}$(R,\phi)(f_\phi)$, $R$ is a Euclidean ring with associated Euclidean norm $\phi$ and adequate division and remainder functions given by $f_\phi$. The termination of the algorithm is a 
%MAR I copied the termination TCC below, but it is well-explained and we do not need to add it to the paper. 
%\linkThLLN{ring_euclidean_algorithm.tccs}{105}{119}
	% % Termination TCC generated (at line 44, column 13) for
	% % euclidean_gcd_algorithm(R, phi, f_phi)(b, rem)
	%   % untried
	% euclidean_gcd_algorithm_TCC9: OBLIGATION
	% FORALL (R: (euclidean_domain?[T, +, *, zero, one]),
	%           (phi: [(difference(R, singleton(zero))) -> nat]
	%                | euclidean_pair?[T, +, *, zero](R, phi)),
	%           (f_phi: [[(R), (remove(zero, R))] -> [(R), (R)]]
	%                | euclidean_f_phi?[T, +, *, zero](R, phi)(f_phi)),
	%           a: (R), b: (remove[T](zero, R))):
	%     NOT a = zero AND phi(a) >= phi(b) IMPLIES
	%      FORALL (rem: (R)):
	%        rem = (f_phi(a, b))`2 AND NOT rem = zero IMPLIES
	%         lex2(phi(rem), IF b = zero THEN 0 ELSE phi(b) ENDIF) <
	%          lex2(phi(b), IF a = zero THEN 0 ELSE phi(a) ENDIF)
\textit{proof obligation} (termination TCC) automatically generated by PVS using the lexicographical {\tt MEASURE} given in the specification. This measure decreases after each possible recursive call: for {\tt Euclidean\_gcd\_algorithm}$(R,\phi,f_\phi)(a,b)$, if $a \neq zero$,  $\phi(a) \geq \phi(b)$ and $rem \neq zero$, the recursive call is {\tt Euclidean\_gcd\_algorithm}$(R,\phi,f_\phi)(b,rem)$; thus, the pair $(\phi(b),\phi(a))$ is lexicographically greater than  $(\phi(rem),\phi(b))$, since $\phi(b)> \phi(rem)$. 
In the other case, the recursive call is  {\tt Euclidean\_gcd\_algorithm}$(R,\phi,f_\phi)(b,a)$. This happens if $a \neq zero$, and $\phi(b) > \phi(a)$; therefore, $(\phi(b),\phi(a))$ is lexicographically greater than   $(\phi(a),\phi(b))$.

It is worth mentioning that such termination TCCs are generated automatically by PVS, but in general, as in this case, the mandatory proof must be formalized manually.

The proof of correctness of the recursive algorithm is given as a straightforward corollary of the \linkThalgebra{ring_euclidean_algorithm.pvs}{50}{56}{\tt Euclid\_theorem} (in Specification \ref{code:EuclideanAlgorithm}) that establishes the correctness of each recursive step regarding the abstract definition of \linkThalgebra{ring_gcd_def.pvs}{36}{40}{\gcd} given in Specification \ref{code:ring_gcd}. Essentially, this theorem states that given an adequate Euclidean norm $\phi$ and associated function $f_\phi$, the \gcd\ of a pair $(a,b)$ is equal to the \gcd\ of the pair $(rem, b)$, where $rem$ is computed through $f_\phi$, i.e.,  $rem$ is equal to the second projection of $f_\phi(a,b)$.   Notice that since Euclidean rings allow a variety of Euclidean norms and associated functions (e.g., \cite{hungerford80}, \cite{Fraleigh2003}), the definition of \gcd\ is not specified as a function but as the relation \gcd?.

Finally, the proof of correctness of the abstract Euclidean algorithm is obtained by induction, using the lexicographic {\tt MEASURE} of the algorithm. The theorem \linkThalgebra{ring_euclidean_algorithm.pvs}{58}{65}{\tt Euclidean\_gcd\_alg\_correctness} (in Specification \ref{code:EuclideanAlgorithm}) formalizes this fact. For an input pair $(a,b)$, in the inductive step of the proof, when $\phi(b)>\phi(a)$ and the recursive call swaps the arguments, one  assumes that 
\[{\tt gcd?}(R)(\{b,a\}, {\tt Euclidean\_gcd\_algorithm}(R,\phi,f_\phi)(b,a)),\]
which means that {\tt Euclidean\_gcd\_algorithm}$(R,\phi,f_\phi)(b,a)$ computes correctly the \gcd\ of the pair $(b,a)$.  From this assumption, one concludes that  
\[{\tt gcd?}(R)(\{a,b\}, {\tt Euclidean\_gcd\_algorithm}(R,\phi,f_\phi)(a,b)).\]
 Otherwise, when the recursive call is   {\tt Euclidean\_gcd\_algorithm}$(R,\phi,f_\phi)(b,rem)$, which happens if $\phi(a)\geq\phi(b)$,  then $rem = (f_\phi(a,b))'2$, the second component of $f_\phi(a,b)$; by induction hypothesis one has that 
 \[{\tt gcd?}(R)(\{b,rem\}, 
 {\tt Euclidean\_gcd\_algorithm}(R,\phi,f_\phi)(b,rem)).\]
Finally, by application of {\tt Euclid\_theorem}, one concludes that the abstract general Euclidean algorithm correctly computes a \gcd\ for the pair $(a,b)$.

\begin{lstlisting}[float=*, xleftmargin=2mm, linewidth=15.8cm, label = code:EuclideanAlgorithm , caption = {Sub-theory \linkThalgebra{ring_euclidean_algorithm.pvs}{23}{65}{\tt ring\_euclidean\_algorithm}: abstract \gcd\ Euclidean algorithm for Euclidean rings},  mathescape, frame = single]
Euclidean_gcd_algorithm(R : (Euclidean_domain?[T,+,*,zero,one]), 
                        (phi: [(R - {zero}) -> nat] | Euclidean_pair?(R,phi)),
                        (f_phi: [(R),(R - {zero}) -> [(R),(R)]] | 
                                       Euclidean_f_phi?(R,phi)(f_phi)))
                        (a: (R), b: (R - {zero})) : RECURSIVE (R - {zero}) =
  IF  a = zero THEN b 
  ELSIF  phi(a) >= phi(b) THEN 
      LET rem = (f_phi(a,b))`2 IN 
        IF rem = zero THEN b 
        ELSE Euclidean_gcd_algorithm(R,phi,f_phi)(b,rem)
        ENDIF             
  ELSE  Euclidean_gcd_algorithm(R,phi,f_phi)(b,a)
  ENDIF
MEASURE lex2(phi(b), IF a = zero THEN 0 ELSE phi(a) ENDIF)

Euclid_theorem : LEMMA
  FORALL(R:(Euclidean_domain?[T,+,*,zero,one]), 
        (phi: [(R - {zero}) -> nat] | Euclidean_pair?(R, phi)),
        (f_phi: [(R),(R - {zero}) -> [(R),(R)]] | 
                       Euclidean_f_phi?(R,phi)(f_phi)),
        a: (R), b: (R - {zero}), g : (R - {zero})) :
           gcd?(R)({x : (R) | x = a OR x = b}, g) IFF 
           gcd?(R)({x : (R) | x = (f_phi(a,b))`2 OR x = b}, g) 

Euclidean_gcd_alg_correctness : THEOREM 
  FORALL(R:(Euclidean_domain?[T,+,*,zero,one]), 
         (phi: [(R - {zero}) -> nat] | Euclidean_pair?(R, phi)),
         (f_phi: [(R),(R - {zero}) -> [(R),(R)]] | 
                        Euclidean_f_phi?(R,phi)(f_phi)),
         a: (R), b: (R - {zero}) ) :
      gcd?(R)({x : (R) | x = a OR x = b}, 
              Euclidean_gcd_algorithm(R,phi,f_phi)(a,b))
\end{lstlisting}

\begin{lstlisting}[float=*, xleftmargin=2mm, linewidth=15.8cm, label = code:ring_gcd , caption = {\gcd\ definition for commutative rings - sub-theory \linkThalgebra{ring_gcd_def.pvs}{23}{45}{\tt ring\_gcd\_def}},  mathescape, frame = single]
 gcd?(R)(X: {X | NOT empty?(X) AND subset?(X,R)}, d:(R - {zero})): bool =
     (FORALL a: member(a, X) IMPLIES divides?(R)(d,a)) AND
	  (FORALL (c:(R - {zero})):
	    (FORALL a: member(a, X) IMPLIES divides?(R)(c,a)) IMPLIES 
     divides?(R)(c,d))
\end{lstlisting}

Now, we show how the correctness of the abstract algorithm  {\tt Euclidean\_gcd\_algorithm} is easily inherited, under adequate instantiations, for the structures of integers $\mathbb{Z}$ and Gaussian integers $\mathbb{Z}[i]$.  The lines of reasoning follow those given in discussions on factorization in commutative rings and multiplicative norms in textbooks (e.g., Section 47 in \cite{Fraleigh2003}, or Chapter 3, Section 3 in \cite{hungerford80}).

The Specification \ref{code:gcdCorrectnessZ} presents the case of the Euclidean ring $\mathbb{Z}$.  The Euclidean norm $\phi_{\mathbb{Z}}$ is selected as the absolute value while the associated function $f_{\phi_\mathbb{Z}}$ is built using the integer division and remainder, specified in the PVS prelude libraries as {\tt ndiv} and \rem: for  $a\in\mathbb{Z}, b\in \mathbb{Z}\setminus \{0\}$, \ndiv$(a,b)$ computes the integer division of $a$ by $b$, and, for $b\in \mathbb{Z}^+\setminus \{0\}$, \rem$(b)(a)$ computes the remainder of $a$ by $b$. 

The correctness of the Euclidean algorithm for the ring of integers is specified as the corollary \linkThalgebraex{ring_euclidean_gcd_algorithm_Z.pvs}{41}{43}{\tt Euclidean\_gcd\_alg\_correctness\_in\_Z}. It states that for the Euclidean ring of integers $\mathbb{Z}$, and any $i,j\in \mathbb{Z}, j\neq 0$, the parameterized abstract algorithm, {\tt Euclidean\_gcd\_algorithm[int,+,*,0,1]} satisfies the relation {\tt gcd?[int,+,*,0]}:

\[\begin{array}{l}\gcd?[int,+,*,0](\mathbb{Z})(\{i,j\}, \\
            {\tt Euclidean\_gcd\_algorithm}[int,+,*,0,1] 
            (\mathbb{Z}, \phi_\mathbb{Z},f_{\phi_\mathbb{Z}})(i,j))
            \end{array}\]

The formalization of this corollary follows from the theorem of correctness for the abstract Euclidean algorithm,   {\tt Euclidean\_gcd\_alg\_correctness} theorem (Specification \ref{code:EuclideanAlgorithm}), which essentially requires proving that the chosen Euclidean measure $\phi_\mathbb{Z}$, and the associated function $f_{\phi_\mathbb{Z}}$ fulfill the conditions in the definition of Euclidean rings. The latter is formalized as lemma \linkThalgebraex{ring_euclidean_gcd_algorithm_Z.pvs}{38}{38}{\tt phi\_Z\_and\_f\_phi\_Z\_ok}: ${\tt Euclidean\_f\_phi?[int,+,*,0]}(\mathbb{Z},\phi_\mathbb{Z})(f_{\phi_\mathbb{Z}})$.

\begin{lstlisting}[float=*, xleftmargin=2mm, linewidth=15.8cm, label = code:gcdCorrectnessZ , caption = {Correctness of the parameterization of the abstract Euclidean algorithm for the Euclidean ring $\mathbb{Z}$
-  sub-theory \linkThalgebraex{ring_euclidean_gcd_algorithm_Z.pvs}{24}{45}{\tt ring\_euclidean\_gcd\_algorithm\_Z}},  mathescape, frame = single]
phi_Z(i : int | i /= 0) : posnat =  abs(i)
 
f_phi_Z(i : int, (j : int | j /= 0)) : [int, below[abs(j)]] = 
 ((IF j > 0 THEN ndiv(i,j) ELSE -ndiv(i,-j) ENDIF), rem(abs(j))(i)) 

phi_Z_and_f_phi_Z_ok  : LEMMA Euclidean_f_phi?[int,+,*,0](Z,phi_Z)(f_phi_Z)

Euclidean_gcd_alg_correctness_in_Z  : COROLLARY
  FORALL(i: int, (j: int | j /= 0)  ) :
    gcd?[int,+,*,0](Z)({x : (Z) | x = i OR x = j},
            Euclidean_gcd_algorithm[int,+,*,0,1](Z, phi_Z,f_phi_Z)(i,j))
\end{lstlisting}

The Specification \ref{code:gcdCorrectnessZi} presents the formalization of the correctness of the Euclidean algorithm for the Euclidean ring $\mathbb{Z}[i]$ of Gaussian integers. The Euclidean norm  of a Gaussian integer $x= ({\tt Re}(x) + i\, {\tt Im}(x)) \in\mathbb{Z}[i]$ is considered as the natural number given by 
$\phi_{\mathbb{Z}[i]}(x) = x * {\tt conjugate}(x)$ = ${\tt Re}(x)^2 + {\tt Im}(x)^2$, where
${\tt conjugate}(x) = {\tt Re}(x) - i\, {\tt Im}(x)$. The construction of an adequate associated function $f_{\phi_{\mathbb{Z}[i]}}$ ({\tt f\_phi\_Zi} in Specification \ref{code:gcdCorrectnessZi}) requires additional explanations and is specified through the auxiliary function \linkThalgebraex{ring_euclidean_gcd_algorithm_Zi.pvs}{63}{66}{\tt div\_rem\_appx}. For a pair of integers $(a,b)$, $b\neq 0$, this function computes the pair of integers $(q, r)$ such that $a = q\, b + r$, and $|r| \leq |b|/2$; thus, $q\,b$ is the integer closest to $a$. The equality $a = q\, b + r$ is formalized as lemma \linkThalgebraex{ring_euclidean_gcd_algorithm_Zi.pvs}{69}{72}{\tt div\_rev\_appx\_correctness}. Several properties about the field of complex numbers are imported from the PVS {\tt complex} theory. 

Now, we explain the construction of the function \linkThalgebraex{ring_euclidean_gcd_algorithm_Zi.pvs}{79}{81}{$f_{\phi_{\mathbb{Z}[i]}}$}. For $y$, a Gaussian integer  and  $x$, a positive integer, let ${\tt Re}(y) = q_1 x + r_1$ and ${\tt Im}(y) = q_2 x + r_2$, where $(q_1,r_1)$ and $(q_2,r_2)$ are computed with the auxiliary function  {\tt div\_rem\_appx} (with respective inputs $({\tt Re}(y),x)$ and $({\tt Im}(y),x)$).  Let $ q = q_1 + i q_2$ and $r = r_1 + i r_2$, then $y = q x + r$. Also, notice that if $r \neq 0$ then $\phi_{\mathbb{Z}[i]}(r) \leq \phi_{\mathbb{Z}[i]}(x)$, since $r_1^2 + r_2^2 \leq x^2/2\leq x^2$.  For the case in which $x$ is a nonzero Gaussian integer, $\phi_{\mathbb{Z}[i]}(x) > 0$ holds. 

Then, we can compute  $\mbox{{\tt div\_rem\_appx}$(y\, {\tt conjugate}(x), x\, {\tt conjugate}(x))$,}$  obtaining $q,r'\in \mathbb{Z}[i]$ such that 
$y\, {\tt conjugate}(x) = q\, (x \,{\tt conjugate}(x)) + r',$ and $r'=0$ or $\phi_{\mathbb{Z}[i]}(r') <  \phi_{\mathbb{Z}[i]}(x\,  {\tt conjugate}(x))$. 

By selecting $r = y - q\,x$, we obtain  
$y = q\, x + r \mbox{ and } r\,{\tt conjugate}(x) = r'$. 

Finally, when $r\neq 0$, since  
$\phi_{\mathbb{Z}[i]}(r\,{\tt conjugate}(x)) <  \phi_{\mathbb{Z}[i]}(x\,  {\tt conjugate}(x))$, 
by application of the lemma \linkThalgebraex{ring_euclidean_gcd_algorithm_Zi.pvs}{54}{55}{\tt phi\_Zi\_is\_multiplicative}, we conclude that $\phi_{\mathbb{Z}[i]}(r) < \phi_{\mathbb{Z}[i]}(x)$. 

The formalization of the correctness of the Euclidean algorithm for Gaussian integers obtained by parameterizations with $\mathbb{Z}[i]$, its Euclidean norm $\phi_{\mathbb{Z}[i]}$ and associated function $f_{\phi_{\mathbb{Z}[i]}}$ follows as the simple corollary \linkThalgebraex{ring_euclidean_gcd_algorithm_Zi.pvs}{91}{93}{{\tt Euclidean\_gcd\_alg\_} {\tt in\_Zi}} in Specification \ref{code:gcdCorrectnessZi}. This is proved using the correctness of the abstract Euclidean algorithm (Specification \ref{code:EuclideanAlgorithm}) and lemma \linkThalgebraex{ring_euclidean_gcd_algorithm_Zi.pvs}{84}{84}{\tt phi\_Zi\_and\_f\_phi\_Zi\_ok}. 
The latter states that  the Euclidean norm $\phi_{\mathbb{Z}[i]}$ and its associated function $f_{\phi_{\mathbb{Z}[i]}}$ are adequate for the Euclidean ring  $\mathbb{Z}[i]$:
${\tt Euclidean\_f\_phi?[complex,+,*,0]}(\mathbb{Z}[i],\phi_{\mathbb{Z}[i]})(f_{\phi_{\mathbb{Z}[i]}})$.

\enlargethispage{5mm}
\begin{lstlisting}[float=*, xleftmargin=2mm, linewidth=15.8cm, label = code:gcdCorrectnessZi , caption = {Correctness of the parameterization of the abstract Euclidean algorithm for $\mathbb{Z}\protect{[i]}$ - sub-theory \linkThalgebraex{ring_euclidean_gcd_algorithm_Zi.pvs}{24}{95}{\tt ring\_euclidean\_gcd\_algorithm\_Zi}},  mathescape, frame = single]
Zi: set[complex] = {z : complex | EXISTS (a,b:int): a = Re(z) AND b = Im(z)}

Zi_is_ring: LEMMA ring?[complex,+,*,0](Zi)

Zi_is_integral_domain_w_one: LEMMA integral_domain_w_one?[complex,+,*,0,1](Zi)

phi_Zi(x:(Zi) | x /= 0): nat = x * conjugate(x)

phi_Zi_is_multiplicative: LEMMA
   FORALL((x: (Zi) | x /= 0), (y: (Zi) | y /= 0)): 
                  phi_Zi(x * y) = phi_Zi(x) * phi_Zi(y)
   
div_rem_appx(a: int, (b: int | b /= 0)) : [int, int] =
  LET r = rem(abs(b))(a), 
      q = IF b > 0 THEN ndiv(a,b) ELSE -ndiv(a,-b) ENDIF  IN
   IF r <= abs(b)/2 THEN (q,r) 
   ELSE IF b > 0 THEN (q+1, r - abs(b)) 
        ELSE (q-1, r - abs(b)) 
        ENDIF 
   ENDIF 
   
div_rev_appx_correctness : LEMMA 
   FORALL (a: int, (b: int | b /= 0)) : 
      abs(div_rem_appx(a,b)`2) <= abs(b)/2 AND 
      a = b * div_rem_appx(a,b)`1 +  div_rem_appx(a,b)`2
      
f_phi_Zi(y: (Zi), (x: (Zi) | x /= 0)): [(Zi),(Zi)] =
  LET q = div_rem_appx(Re(y * conjugate(x)), x * conjugate(x))`1 +  
          div_rem_appx(Im(y * conjugate(x)), x * conjugate(x))`1 * i,
      r = y - q * x IN (q,r)

 phi_Zi_and_f_phi_Zi_ok: LEMMA
    Euclidean_f_phi?[complex,+,*,0](Zi,phi_Zi)(f_phi_Zi)

 Euclidean_gcd_alg_in_Zi: COROLLARY
  FORALL(x: (Zi), (y: (Zi) | y /= 0)  ) :
      gcd?[complex,+,*,0](Zi)({z :(Zi) | z = x OR z = y},
       Euclidean_gcd_algorithm[complex,+,*,0,1](Zi, phi_Zi,f_phi_Zi)(x,y))
\end{lstlisting}

%%%%%%%%%%%%%%%%%%%%%%%%%%%%%%%  
\section{Related Work and work in progress} \label{sec:related_work}
%%%%%%%%%%%%%%%%%%%%%%%%%%%%%%%  

\subsection{Related work}

Several formalizations focus on specific ring structures as the ring of integers. Such developments range from simple formalization exercises, such as correctness proofs of \gcd\ algorithms for $\mathbb{Z}$, to elaborated mechanical proofs of the Chinese Remainder theorem for $\mathbb{Z}$.
The latter started from Zhang and Hua's RRL (Rewrite Rule Laboratory) mechanization \cite{ZhangHua92},  followed by different approaches in Mizar, HOL Light, hol98, Coq \cite{Schwarzweller2009},  ACL2 \cite{Russinoff2000}, and VeriFun \cite{Walther2018}.
Also, in the theory {\tt ints@gcd.pvs} from NASA PVS library, one finds Euclid's algorithm restricted to integer numbers.
Nevertheless, the general algebraic abstract approach is followed by a few developments. In particular, such an approach is followed in the Isabelle/HOL Algebra Library (see \cite{AransayD16}, \cite{ballarinizabelle2019}, and \cite{Ballarin2019}); a library that provides a wide range of theorems on mathematical structures, including results on rings, groups, factorization over ideals, rings of integers and polynomial rings, as well as formalization of an algorithm to compute echelon forms over Euclidean domains, and so characteristic polynomials of matrices. Also, the Lean mathlib library \cite{LeanLibrary2020}   specifies unique factorization domains,  prime and irreducible elements in commutative rings,  and relations with principal ideal domains. In addition, it specifies the notion of \gcd\ for Euclidean domains and formalizes several properties as the correctness of the extended Euclidean algorithm by applying Bézout's \gcd\ lemma. The library mathlib formalizes that a Euclidean domain is a principal ideal domain, and a principal ideal domain is a unique factorization domain. The former is given as formally verified construction from a definition. From this instance, it is possible to infer that the Gaussian integers are a Euclidean domain and thus a principal ideal. Also, the Euclidean algorithm can be adapted to structures such as the Gaussian integers. A recent extension of mathlib specifies the ring of Witt vectors and formalizes the isomorphism between the ring of Witt vectors over $\mathbb{Z}/p\mathbb{Z}$ and the ring of $p$-adic integers $\mathbb{Z}p$, for a prime $p$ \cite{CommelinL21}.

In Coq, results about groups, rings, and ordered fields were formalized as part of the FTA project \cite{geuverscoq2002}; this work gave rise to the formalization of the Feit and Thompson's proof of the Odd Order Theorem \cite{Gonthier13}. Also, there are formalizations in Coq of real ordered fields \cite{Cohen2012},  finite fields \cite{Philipoom2018},  and rings with explicit divisibility \cite{cohencoq16}. In Nuprl and Mizar, there are proofs of the Binomial Theorem for rings in \cite{jackson1995} and \cite{cs01}, respectively, and a Mizar formalization of the First Isomorphism Theorem for rings \cite{Kornilowicz2014}. ACL2 has a hierarchy of algebraic structures ranging from setoids to vector spaces that aims the formalization of computer algebra systems \cite{heras15}. 

Regarding the paper-and-pen proofs in \cite{hungerford80} and the formalization reported in this paper, the last one comprises about twenty pages of Hungerford's textbook. We estimate it took ten months of human labor. Some results in the book appear as trivial remarks only. Nevertheless, they required the formalization of a significative sequence of auxiliary lemmas. An excellent example of the lack of details in this respect is a remark after Definition 3.5. in \cite{hungerford80} that we have used in tutorials to motivate mathematicians to deal with proof assistants. It states that: 

\begin{quote}
    \textit{``every irreducible element in a unique factorization domain is necessarily prime by Definition 3.5. Consequently, irreducible and prime elements coincide, by Theorem 3.4.''}
\end{quote}  

Indeed, the formalization of this remark required the application of additional properties related to bijective functions, the equivalence relation ``associates'', and the composition of finite sequences, among others inherited from the abstract structure integral domain. 
Also, several particular cases had to be analyzed to 
ensure the result when the elements involved are units or equal to zero. 

Finally, we would like to stress that the project is focused on the formalization side, but aspects related to code extraction can be explored through tools provided by PVS. For instance, PVSio is an animation tool that extends the ground evaluator of PVS with a predefined library of programming features \cite{PVSio2003}. The evaluation is possible whenever algorithms are specified constructively. For our purposes, this means that we can run the formalized \gcd\ algorithm with the help of the ground evaluator for any Euclidean domain for which the Euclidean norm $\phi$ and the associated function $f_\phi$ are specified constructively, which is the case of our specifications for $\mathbb{Z}$ and $\mathbb{Z}[i]$. Based on PVSio, elaborated approaches use this animation tool to evaluate the formal models on a set of randomly generated test cases, comparing the computed results against output values obtained by actual software \cite{DutleMNB15}. After applying such approaches, performance and comparison with other implementations would be possible.

\subsection{Work in progress and applications}

Work in progress to be reported in the future includes formalizing the general theory of quaternions built from any abstract structure of fields specified in PVS as commutative division rings. The specification of quaternions is given from an abstract type {\tt T} with binary operators for addition and multiplication, with constants {\tt zero} and {\tt one}, respectively. The type {\tt T} with addition and {\tt zero} is an Abelian group, and the multiplication is associative. The specification includes axioms for quaternion addition and multiplication ($i^2 = a$, $j^2 = b$, for some given parameters $a$ and $b$ of $T$), associativity for quaternion multiplication, distributivity of quaternion addition and multiplication, and properties for the scalar product between elements of the field and of the quaternion. All that is provided in the theory \linkThalgebra{quaternions_def.pvs}{22}{93}{quaternion\_def}.  Afterward, in the PVS theory \linkThalgebra{quaternions.pvs}{22}{210}{quaternions}, using these axioms, a series of general properties of quaternions are provided, which range from the characterization of quaternion multiplication to the characterization of quaternions as division rings. Once again, following the general approach to specifying quaternions from abstract fields, we can obtain the specific structure of Hamilton's quaternions as the theory \linkThalgebra{quaternions_Hamilton.pvs}{24}{178}{quaternions\_Hamilton}, using as the parameter to build the quaternions the specific field of reals \cite{AyalaRinconITLPAR2023}. As far as we know, there are formalizations of Hamilton's quaternions in HOL Light and Isabelle/HOL (e.g., \cite{GaMa2017}, \cite{Paulson18}). In contrast, some elements of the general theory of quaternions built over any abstract field, as in our case, were developed as part of the Lean mathlib library \cite{LeanLibrary2020}.

We do not argue that any proof assistant is a better or worse framework than any other for formalizing algebraic notions and properties. However, we are confident that the current formalization work adequately explores the inductive and higher-order possibilities available in PVS and substantially contributes to completing the theory of the algebraic properties of rings by providing the most general and abstract possible presentation of such algebraic structures, as also given in some of the previous references, mainly as done by the approaches mentioned above in Isabelle/HOL and Lean (\cite{ballarinizabelle2019}, \cite{LeanLibrary2020}). 
%%%%%%%%%%%%%%%%%%%%%%%%%%%%%%%  
%\subsection{Formalization vs pen-and-paper proofs}

%%%%%%%%%%%%%%%%%%%%%%%%%%%%%%%  
\section{Conclusions and Future Work} \label{sec:conclusion}

In contrast to other works, restricted to specific ring structures, our formalization approach focuses on the theory of abstract rings, as done in the Lean- and Isabelle-related libraries (cf \cite{LeanLibrary2020} and \cite{Ballarin2019}, respectively) discussed in the related work. Advantages of such an approach include increasing the interest of mathematicians in formalizations and having practical general presentations of computational algebraic properties portable to specific ring structures. In particular, in \cite{LimaGAA21}, the Chinese Remainder Theorem was formalized for (non-necessarily commutative) rings, obtaining, as a corollary, the CRT version for the ring of integers. This work substantially extends the  {\tt algebra} PVS library by specifying  Euclidean rings and factorization domains and formalizing the correspondence between principal ideal domains and unique factorization domains. Also, it proves the correctness of a general Euclidean \gcd\ algorithm for Euclidean domains. The usefulness of such an abstract verified \gcd\ algorithm is evident by its adaptation to specific Euclidean domain structures. Indeed, this versatility is illustrated by showing how simple corollaries establish the correctness of the Euclidean algorithm (parameterized) for the rings of integers and Gaussian integers ($\mathbb{Z}$ and $\mathbb{Z}[i]$).   

In future work, we will include the specification of modular arithmetic and verification of generic versions of Euler's Theorem and Fermat's Little Theorem for Euclidean domains.

%\href{https://isabelle.in.tum.de/library/HOL/HOL-Analysis/HOL-Computational_Algebra.Factorial_Ring.html}{https://isabelle.in.tum.de/library/HOL/HOL-Analysis/HOL-Computational\_Algebra.Factorial\_Ring.html}

%\href{https://isabelle.in.tum.de/library/HOL/HOL/GCD.html}{https://isabelle.in.tum.de/library/HOL/HOL/GCD.html}

%\href{https://www.isa-afp.org/entries/Echelon_Form.html}{https://www.isa-afp.org/entries/Echelon\_Form.html}

%\href{https://www.isa-afp.org/theories/echelon_form/#Rings2}{https://www.isa-afp.org/theories/echelon\_form/#Rings2}

%\href{https://dl.acm.org/doi/10.1007/s00165-016-0383-1}{https://dl.acm.org/doi/10.1007/s00165-016-0383-1}

\bibliographystyle{eptcs}
%\bibliography{JAR_final_submission/bibliography}

\end{document}